\documentclass[a4paper,11pt]{article}
\usepackage[pdftex]{graphicx}
\usepackage{amsmath,amsfonts}

\newcommand{\vett}[1]{\mathbf{#1}}

\newcommand{\diff}{\mathrm{d}}
\newcommand{\veps}{  {\varepsilon} }
\newcommand{\interi}{\mathinner{\bf Z}}

\newcommand{\udf}{ \stackrel{\mathrm{def}}{=} }

\makeatletter
    \renewcommand*{\@fnsymbol}[1]{\ensuremath{\ifcase#1\or
 \textrm{a}\or
        \textrm{b}\or \textrm{c}\or \mathsection\or \mathparagraph\or
 \|\or
        **\or \dagger\dagger  \or \ddagger\ddagger \else\@ctrerr\fi}}
\makeatother

\title{Tsallis distributions, their relaxations  and the relation
  $\Delta t \cdot \Delta E \simeq h$, in the dynamical fluctuations of a
  classical model of a crystal}

\author{
  A. Carati\thanks{ Dipartimento di Matematica, Universit\`a di
    Milano, Via Saldini 50, 20133 Milano, Italy. E-mail:
    {\tt andrea.carati@unimi.it} }
  \and L. Galgani\footnotemark[1]
  \and F. Gangemi\thanks{
    DMMT, Universit\`a di Brescia, Viale Europa 11, 25123 Brescia,
    Italy.} 
  \and
  R. Gangemi\footnotemark[2] }

\date{\today}

\begin{document}

\maketitle


\begin{abstract}
  We report the results of a numerical investigation, performed in the
  frame of dynamical systems' theory, for a realistic model of a ionic
  crystal for which, due to the presence of long--range Coulomb
  interactions, the Gibbs distribution is not well defined.  Taking
  initial data with a Maxwell-Boltzmann distribution for the
  mode-energies $E_k$, we study the dynamical fluctuations, computing
  the moduli of the the energy-changes $|E_k(t)-E_k(0)|$.  The main
  result is that they follow Tsallis distributions, which relax to
  distributions close to Maxwell-Boltzmann ones; indications are also
  given that the system remains correlated.  The relaxation time
  $\tau$ depends on specific energy $\veps$, and for the curve $\tau$
  vs, $\veps$ one has two results. First, there exists an energy
  threshold $\veps_0$, above which the curve has the form
  \begin{equation*}
  \tau \cdot \veps \simeq h\ ,
  \end{equation*}
  where, unexpectedly, Planck's constant $h$ shows up.  In terms of
  the standard deviation $\Delta E$ of a mode-energy (for which one
  has $\Delta E=\veps$), denoting by $\Delta t$ the relaxation time
  $\tau$, the relation reads $\Delta t \cdot \Delta E \simeq h$, which
  reminds of the Heisenberg uncertainty relation. Moreover, the
  threshold corresponds to zero-point energy. Indeed, the quantum
  value of the latter is $h\nu/2$ ( where $\nu$ is the characterisic
  infrared frequency of the system), while we find $\veps_0 \simeq
  h\nu/4$, so that one only has a discrepancy of a factor 2.
  So it seems that lack of full chaoticity manifests itself, in
  Statistical Thermodynamics, through quantum-like phenomena.
\end{abstract}

\section{Introduction. A modified version of the original FPU problem}

As pointed out by several authors,\footnote{The first one known to the
  present authors is Khinchin in his book ref.~\cite{khinchin},
  Section 20.}  simple large deviation arguments of Boltzmann type
show that the distribution of the normal-mode energies of a system of
weakly coupled oscillators is generically a Maxwell--Boltzmann one. In
other terms, the region of a constant-energy surface presenting a
distribution different from Maxwell-Boltzmann, has negligible measure.

This however is not sufficient for the aims of Statistical
Thermodynamics, since many quantities of interest are expressed in
terms of correlations of suitable observables (see
\cite{CGM2017}), computed from initial data of generic
type (i.e., with a Maxwell-Boltzmann distribution), and not just in term
of averages of the normal-mode energies.  An example is
given in the works \cite{CGMGG2015,CGMGG2018}, where the electric susceptibilty
of Lithium Fluoride (LiF) was computed through the Fourier transform
of the auto-correlation of the optical normal-mode amplitudes, for
generic initial data, and the infrared spectra thus obtained were
found to agree very well with the experimental ones at several
temperatures (more details will be given later).

In other words, the dynamical properties of the considered system are
relevant, beyond the occurrence of Maxwell-Boltzmann distributions.
So, leaving aside questions of ergodic type (see
\cite{flach17,flach18,kozlov02}), we decided to investigate whether
significant properties of some general character may be found to occur
for the dynamical fluctuations in systems of coupled oscillators,
still working on the Lithium Fluoride model, which allows a comparison
to be made between computational results and experimental data.

The problem is then to choose the significant quantities to be
investigated, and our choice was to study the probability
distributions for the moduli $ | E_k(t)- E_k(0) |$ of the
energy-changes of the normal modes.  This was inspired to the
following naive argument.\footnote{Such an  argument was
  conceived by the first author about ten years ago, and some
  preliminary unpublished results for a standard Fermi-Pasta-Ulam
  model were illustrated in the thesis~\cite{dubrovich} of
  M. Dubrovich, with the first two authors as advisors.  The results
  concerned the energy-changes of single normal modes, whose
  distribution was found to be depend on mode-frequency. Tsallis
  distributions were not taken into consideration, and no comparison
  with experimental results was possible, so that no interpretation of
  the results could be advanced.}
Assume that, starting from a point $x_0$ in
phase space, the evolved point $x_1 \stackrel{\rm def}{=} \Phi^t x_0$
at time $t$ may be considered as ``independent'' of $x_0$, in the
sense that the normal-mode energies may be considered as random
variables independent of their initial values.  Then it will turn out
that a Maxwell-Boltzmann distribution occurs not only for the
normal-mode energies, both in $x_0$ and in $x_1$, but also (as a
simple calculation shows - see Section 3) for the moduli of the
energy-changes $ | E_k(t)- E_k(0) |$. By an inversion of the argument,
our choice was to try to establish whether the distribution of such
moduli of the the energy-changes converges to a Maxwell-Boltzmann one,
and on which time scale. This actually constitutes what we call
\emph{a modified version of the original FPU problem},
inasmuch as the latter is
concerned with a relaxation of the distribution of the mode-energies
themselves,\footnote{ After
  about fifty years of harsh debates (see for example
  refs.~\cite{gallavotti1,campbell05,CGG2005}) the scientific community
  eventually came to agree that, in models of Fermi- Pasta-Ulam type,
  equipartition, and possibly also a Maxwell-Boltzmann distribution,
  would be attained for the normal-mode energies, even if one starts
  from exceptional initial data.}
rather than of their changes.  Our choice may also be looked upon
in the following way.  As particularly stressed by Nernst
\cite{nernst}, energy changes and not energies themselves, are the
relevant quantities for Statistical Thermodynamics.  Now, in systems
of weakly coupled oscillators, as are the Fermi-Pasta-Ulam ones, the
prototypes of the energy-changes are the mode-energy changes,  whose
moduli lead, by averaging, to correlations. These
are indeed the quantities we decided to investigate.

Some interesting results were found. The first one is that at all
times the moduli of the energy-changes seem to follow a well defined
distribution, i.e., a Tsallis one (see \cite{tsallis1988}).  Such a fact
is not completely unexpected, since Tsallis distributions are known to
show up in models involving long-range  forces, either Coulomb
ones \cite{CGGG2020a} (as the  ionic crystal  considered here),
 or gravitational ones (see \cite{rapisarda01} \cite{ruffo02}).
As is known, Tsallis distributions are characterized by two
parameters: $q$ (the so-called \emph{Tsallis entropic index}, related
to the decay of the tail of the distribution), and $\beta$, which
determines the specific energy per degree of freedom $\veps$.  Such
distributions have the form
\begin{equation}
   \rho = C\big[1 - (1-q)\beta E \big]^{\frac 1{1-q}}\ ,
\end{equation}
$C$ being a normalization factor, and reduce to the Maxwell-Boltzmann
one, $C' \exp (-\beta E)$, in the limit $q\to 1$ (which implies that
$\veps$ tends to $1/\beta$). The dynamical reason for the occurring of
a Tsallis distribution in the frame of dynamical systems is not clear
to us, at the moment. However, as pointed out for example in the work
\cite{C2008}, one might surmise that such an occurrence could
be related to the ``dimension'' of the orbits in phase space, i.e., to
their tendency to invade sets with dimensions smaller than
that of the energy surfaces, as if some remembrance remained of the
integrals of motion of the linearized system.

In our computations the parameters $q$ and $\beta$ were found to
change with time, presenting some final relaxation for large enough
times.  At first sight, the relaxation might appear to lead to a
Maxwell-Boltzmann distribution, but a more accurate inspection shows
that the results depend on the number $N$ of ions. Thus, for $N=4096$
(and specific energy per degree of freedom $\veps=500$
Kelvin\footnote{All along the present work, the specific energies are
  expressed in Kelvin, i.e., as their ratios with respect to the
  Boltzmann constant $k_B$. This will be seen to make a comparison
  with the experimental data easier.}), the relaxed distribution turns
out not to coincide with Maxwell-Boltzmann, being however rather close
to it, since one finds $q=1.014$. This  indicates that some kind
of correlation persists even after relaxation.

Then, the dependence of the relaxation time $\tau$ on the specific
energy $\veps$ per degree of freedom was investigated.  Here, two
further significant results were found. First of all, there exists a
specific energy threshold $\veps_0$, at about $100$ K, distinguishing
between two behaviors: a quick relaxation above threshold, and a
slower one below it. An analogous behavior was observed long ago for
the Fermi-Pasta-Ulam model, first by Bocchieri, Scotti and Loinger
\cite{bocchieri1970}, then by Casartelli et al. \cite{casartelli1976}
as a vanishing of the maximal Lyapunov exponent, and eventually, by
Pettini and Landolfi in the work \cite{pettini1990}, still in terms of
the Lyapunov exponent in a way very similar to the one found here,
actually as a knee in the curve Lyapunov exponent versus specific
energy. Concerning such a threshold, already in the paper
\cite{CGS1972}, along the lines of \cite{GS1972}, the idea was
advanced by Carlo Cercignani that it should correspond to zero-point
energy.  This is indeed essentially confirmed here, since it turns out
(see later) that, denoting by $\nu$ the characteristic infrared
frequency of Lithium Fluoride, for the threshold energy $\veps_0$ one
has $h\nu/2 \simeq 2\veps_0$, where $h$ is Planck's constant, so that,
up to a factor 2, $\veps_0$ coincides with the quantum zero-point
energy.
 
Finally, above threshold another result was met, and again a rather peculiar
one. Indeed the relaxation time was found to be inversely
proportional to specific energy, through a proportionality constant, an action,
which again  turns out to be nearly the Planck constant (actually, about
three times larger), so that one has\footnote{More precisely, one has
  $\tau(\veps)\simeq 3.5 \, h/\veps$ so that, with $\veps_0\simeq
  h\nu/4$, for the largest relaxation time $\tau_{max}\simeq 3.5\,
  h/\veps_0$ one finds $\tau_{max}\simeq 1/\nu$, i.e., the maximal
  relaxation time is equal to the period of the characteristic
  infrared frequency.}
\begin{equation}\label{tempi}
  \tau(\veps)\simeq \frac h\veps \quad\quad \mathrm{i.e.,} \quad\quad
  \tau \cdot \veps \simeq h \ .
\end{equation}

Now, every mode-energy is always Maxwell-Boltzmann distributed, so
that for its standard deviation $\Delta E$ one has
\begin{equation}\label{deltaE}
  \Delta E = \veps \ .
\end{equation}
Thus, the result (\ref{tempi}) for the relaxation time can be phrased
in the following suggestive way: \emph{The product of the
  uncertainty of a mode--energy by the relaxation time is nearly equal to $h$},
\begin{equation}\label{heis}
\tau \cdot \Delta E \simeq h \ .
\end{equation}
which closely reminds of the Heisenberg uncertainty principle.

We are well aware  that the problems on the foundations of Statistical
Thermodynamics are very difficult and complex. However
it seems to us that  the above mentioned results, although obtained in a quite
elementary way, may present  a certain interest.

\vspace{1 em}

In Section~2 the Lithium Fluoride model is briefly described, and are
recalled the relevant previous results on electric susceptibility
which motivated the present study. In particular it is recalled how
the results on the infrared spectra naturally make zero-point energy
emerge in a classical frame. Section~3 is instead devoted to
illustrating the new numerical results.  In Section 4 the results
obtained are summarized and briefly discussed, commenting in
particular how should one understand the occurring of Planck's
constant in a purely classical model, while some conclusions are drawn
in the last Section.

\section{The model, the infrared spectra, and the idea of  a classical
way to zero-point energy}

We briefly recall the ionic-crystal model investigated, and the
relevant results available in the literature.  In the studies on the
dynamics of the ions in an ionic crystal, following Born one usually
works in the so-called adiabatic approximation, in which the dynamics
of the electrons can be ignored, and the ions are dealt with as point
particles. Indeed, the screening effect of the electrons is implicitly
taken into account both through a suitable interaction potential
acting among the ions, and through an effective charge for the ions in
place of the true one.  In such an approximation the following model
can be introduced. One considers a system of an even number $N$ of
point particles $\vett x_{i,s}$, $i=1,\ldots,N/2$ of two suitable
ionic species $s=1,2$ such as Lithium and Fluorine (with masses $m_s$
and charges $e^{(s)}$), located inside a cubic cell of side
$L$.\footnote{At variance with ref.~\cite{CGMGG2015}, in which we worked at fixed
  pressure, now we work at fixed volume
  at all energies, with a lattice step $L= 1.995$
  \AA.} Such points interact both through Coulomb forces, and
through a Buckingham potential of the form\footnote{This is a modern
  modification of the potential $V(r)=C/r^6$, first proposed by Born,
  which contains just one free parameter.  This potential was used in
  our first paper on LiF, and indeed it too turned out to produce
  spectra in good aagreement with experiment, although not as good as
  with the Buckingham potential.}
\begin{equation}
 V_{ss'}(r)=a_{ss'}e^{-b_{ss'}r} + \frac {c_{ss'}}{r^6} \ .
\end{equation}
The values of the constants $a_{ss'}$, $b_{ss'}$ and $c_{ss'}$ and of
the effective charges can be found in the
work \cite{CGMGG2015}, Table I, and were determined by optimizing the
dispersion relations and the spectra at 300 K. Finally, as usual,
periodic boundary conditions are introduced, so that the model turns
out to be defined by the Hamiltonian
\begin{equation}\label{eq:2}
H=\sum_{j,s } \frac {p_{j,s}^2}{2m_s} + \sum_{\vett n\in \interi^3}
\sum_{j,j',s,s'} \Big[ V_{ss'}\big(| \vett x_{j,s} - \vett x_{j',s'} +
  \vett nL|\big) + \frac{e^{(s)}_{eff}e^{(s')}_{eff}}{| \vett x_{i,s}
    - \vett x_{j,s'} + \vett nL|} \Big] \ .
\end{equation}
Actually, such a definition of the model only has a formal character,
inasmuch as the series involving the Coulomb potential is not
absolutely convergent. The definition is completed by prescribing that
such a series be summed according to the Ewald method, which
transforms it into the sum of two rapidly absolutely convergent
series, of which one is defined over the direct lattice, and the other
one over the reciprocal lattice. For the Ewald method one can consult,
in addition to the orignal german work \cite{ewald1921}, also the recent
one \cite{gibbon02}, or our work \cite{CGMBZ2014}.

The first interesting point is that such a Hamiltonian admits a stable
equilibrium configuration (which is chosen as defining the energy
minimum), in which the ionic positions $\vett x_{j,s}^0$ form a
face-centered cubic lattice, which indeed is the structure of LiF.
Furthermore, for not too large kinetic energies (say, below
1060~K),\footnote{The melting temperature is 1118 K.} the ions are
found to oscillate about such equilibrium positions, thus reproducing
the oscillating crystal structure of the solid.

Let us now come to the infrared spectrum. In this connection the
motion of the electrons can be essentially ignored, since it is known
to contribute to susceptibility (in such a domain) only through a
constant. Thus, having obtained, through numerical computations, the
solutions of the equations of motion of the ions corresponding to the
Hamiltonian (\ref{eq:2}), the dielectric susceptibility tensor
\footnote{We recall that  the dielectric tensor
  $\epsilon_{i,j}$ is a thermodynamic quantity, and
  is equal to the Hessian of the Gibbs
  free energy of the crystal with respect to the electric field
  intensity, i.e., one has  $\epsilon_{i,j}=\frac
  {\partial^2 G}{\partial E_i\partial E_j }$ (see
  \cite{grindlay}). } $\chi_{ik}(\omega)$ due to the ions is
computed in the following way. One defines the (microscopic)
polarization as
\begin{equation}
\vett P(t) \udf \frac 1V \sum_{j,s} e^{(s)}_j \big( \vett x_{j,s}(t) -
\vett x_{j,s}^0\big) \   
\end{equation}
with $V=L^3$,
and the susceptibilty is then given by the (Green--Kubo) formula (see
\cite{CG2014})
\begin{equation}
\chi_{jk}(\omega)= \frac 1{\sigma}\int_0^{+\infty} \diff t \,
e^{i\omega t} <P_j(t)\dot P_k(0) > \ ,
\end{equation}
where $\sigma^2$ is the variance of the kinetic energy, while by $<~>$
we have denoted a phase average, which concretely was computed as the
mean of time--averages  over several trajectories.  Thus, in words,
the electric susceptibility is computed essentially as the Fourier transform of
the  time auto-correlation of polarization. In the terminology of dynamical
systems, it is computed in terms of the spectrum of the considered
dynamical system.

In order to perform a comparison with the experimental data, one is
confronted with the problem of identifying temperature, inasmuch as
the experimental measures are performed at a given  temperature, whereas
in our model the computations are performed at a given specific energy.
If one makes use of the familiar identification of temperature as
proportional to mean kinetic energy, the susceptibility thus computed
is found to be  in remarkable agreement with the experimental measures at room
temperature and at larger ones, up to 1000 K.  Instead, the agreement is not so
good at 85 K, and a complete disagreement is found at 7.5 K.

The unexpected empirical result of the work \cite{CGGG2019} was that an
agreement between experimental data and numerical computations at low
temperatures can be restored, provided one abandons the familiar
identification of temperature as proportional to mean kinetic energy.
Indeed the experimental data at a temperature of 85 K are reproduced
if one takes a specific energy $\veps=180$ K, and analogously for the
data at 7.5 K if one takes $\veps =125$ K. One should notice that a
procedure of such a type for constructing the relation between
temperature and mechanical energy has a noble precursor in statistical
mechanics since, in the case of dilute gases, Clausius obtained such a
relation by comparing the theoretical expression of the product $pV$
in terms of kinetic energy, with the empirical expression  in terms
of temperature.\footnote{ An analogous procedure could perhaps be used
  for crystals, by numerically computing pressure as a function of
  energy, and this indeed is an interesting subject for future work.}

With respect to the case of dilute gases, in the case of crystals the
qualitative difference is that a zero-point energy is empirically
known to exist, namely, a finite kinetic energy of the ions is
observed even for vanishing temperature, as proved by the Debye-Waller
effect. Now, the existence of a zero-point energy in crystals is
predicted by quantum mechanics, and one might be induced to think that
at low temperatures one is entering an essentially quantum regime,
irreducible to a classical one.  On the other hand it is a fact that,
through a suitable redefinition of temperature in mechanical terms,
classical dynamics does reproduce the experimental spectra in an
appropriate manner. Thus the problem seems rather to be one of
statistical mechanics, namely, how should temperature be consistently
defined for our system.  This would be trivial if one could make use
of the Gibbs measure. However, in our model we are in presence of long
range potentials (even resummed through the Ewald procedure), so that
the existence of the Gibbs measure is not even
guaranteed. Furthermore, the use of the Gibbs measure would require
that suitable ergodicity conditions be satisfied.

So it seems that the problem of defining temperature in classical
statistical mechanics for a crystal, in a way consistent with
dynamics, is still open. This is indeed the original motivation of the
present work, namely, to investigate the statistical properties
induced by the dynamics, at several values of specific mechanical
energy, for generic initial data. The aim was to check whether some
sort of qualitative change in the dynamics  may be found to occur for specific
energies in the region where, according to the mentioned
results on infrared spectra, zero-point energy should emerge, i.e.,
for specific energies of about 120 K.  This indeed might advocate the
necessity of identifying temperature, in mechanical terms, in a way
different than the Clausius one, in that region of specific energy,
for a LiF crystal. We will show in the next section that this is
actually the case.

\section{The Tsallis distributions for the  moduli of the energy-changes, and their
  relaxation}

To investigate the ergodicity properties of a system with a very large
number of degrees of freedom is not a simple task since, due to the
large deviation theorems, extremely numerous are the functions which
are essentially constant on any given energy surface (i.e., coincide,
apart from regions of negligible measure, with their phase average).
This is essentially the same property that makes the Maxwell-Boltzmann
distribution have an overwhelming probability.  A positive use of this
fact can be made, by trying to check whether a distribution different
from the Maxwell-Boltzmann one can be observed for suitable variables,
since this would
indicate that the dynamics of the considered system presents peculiar
properties.

In such a spirit we decided to look at the mode-energy-changes,
defined in the following way.  As explained in the previous section,
the Hamiltonian admits a configuration of stable
equilibrium,\footnote{Actually the system admits $\simeq N!/2^N$ such
  configurations, which are obtained by permutations of particles of
  the same species.  However, as already mentioned, starting near to
  one of such configuration the point will remain near to it for the
  whole actual integration time.  Different minima can be visited only
  for temperatures above tre melting one.  Then the ions freely
  exhange positions, but the system behaves as a liquid, and no more
  as a solid.} and let the corresponding positions of the ions be
denoted by $\vett x_{j,s}^0$. By linearization about the equilibrium
configuration one obtains a linear system, whose normal mode
amplitudes $A_{\vett k,\rho}$ can be determined, where
$\rho=1,\ldots,6$ denotes one of the 6 branches of the dispersion
relation, and $\vett k$ is a suitable wave vector of the reciprocal
lattice.  The energy of any normal mode is then defined as
\begin{equation}
 E_{\vett k,\rho} = \frac 12\Big( \dot A_{\vett k,\rho}^2 +
 \omega_{\vett k,\rho}^2 A_{\vett k,\rho}^2\Big) \ ,
\end{equation}
where $\omega_{\vett k,\rho}$ is the corresponding oscillation
frequency.  Each $E_{\vett k,\rho}$ is a constant of motion for the
linearized system, whereas it behaves as a ``slow'' variable for the
complete system.

Take now, at random, a phase space point sufficiently near to a given
equilibrium configuration, and consider the set of the corresponding
normal-mode energies $E_{\vett k,\rho}$, computed for such an initial
datum.  As mentioned in the introduction, according to a large
deviation argument it turns out that, if one builds the corresponding
histogram, i.e., one determines how many normal modes have energies in
the interval between $E$ and $E+\delta E$, then the histogram will
follow with a good approximation a Maxwell-Boltzmann law
$C\exp(-E/\veps)$, at least in the limit of large values for the
number $N$ of particles.  In the same way, if one takes two initial
data at random on the same energy surface, still in a neighborhood of
the given equilibrium point, and computes the normal-mode energies
$E_{\vett k,\rho}^0$ and $E_{\vett k,\rho}^1$ corresponding to the
first and the second point respectively, then the histogram of the
moduli of the energy-changes $ \Big | E_{\vett k,\rho}^1 - E_{\vett
  k,\rho}^0\Big |$ will still follow a Maxwell-Boltzmann type law
$C'\exp(-\Delta E/\veps)$, as is easily checked.\footnote{If $x$ and
  $y$ are independent positive random variables, identically
  distributed according the Maxwell-Boltzmann distribution, then the
  probability $P(|x-y|>\Delta E)$ will be
  $$
  P(|x-y|>\Delta E) = \beta^2 \int_{|x-y|>\Delta E}
  e^{-\beta(x+y)}\diff x\,\diff y \ ;
  $$
  performing the change of variable $X=x+y$, $Y=x-y$, the integral
  will change into
  $$
  P(|Y|>\Delta E) = \beta^2 \int_{\Delta E}^{+\infty} e^{-\beta X}
  \diff X \int_{\Delta E}^{X} \diff Y= e^{-\beta \Delta E}
  \int_0^{+\infty} z e^{-z}\diff z = e^{-\beta \Delta E}\ ,
  $$
  having defined $z \udf \beta (X-\Delta E)$.  This shows that
  $|x-y|$ too is distributed according to Maxwell-Boltzmann.  }

Now, instead of taking two points at random, one can take at random just one
point, and take as a second point the evolved one of the first at time $t$,
then  building the histogram
of the moduli of the energy-changes $ \Big | E_{\vett k,\rho}(t) - E_{\vett
  k,\rho}(0)\Big |$: the distribution will obviously be initially, at
$t=0$, a delta centered on 0, and will then change with time. If the
dynamics is sufficiently chaotic, then the evolved point will become
independent of the initial one, so that the distribution may be
expected to converge to a Maxwell-Boltzmann one. Conversely, the
evolved point and the initial one will turn out to be correlated, as
long as the distribution is different from the Maxwell-Boltzmann one.
\begin{figure}
  \begin{center}
    \includegraphics[width=\textwidth]{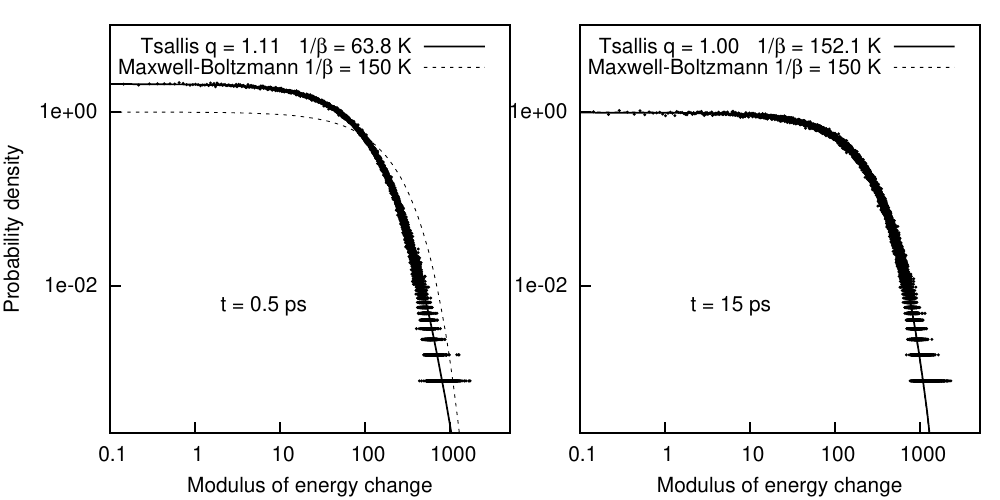}
  \end{center}
  \caption{\label{fig:1} Evolution of the histogram of the moduli of the
    energy-changes, for specific energy $\veps=150$ K and a cell with $N=216$
    ions, as time is
    increased: $t=0.5$ picoseconds left, $t=15$ picoseconds right.}
\end{figure}
\begin{figure}
  \begin{center}
    \includegraphics[width=1.\textwidth]{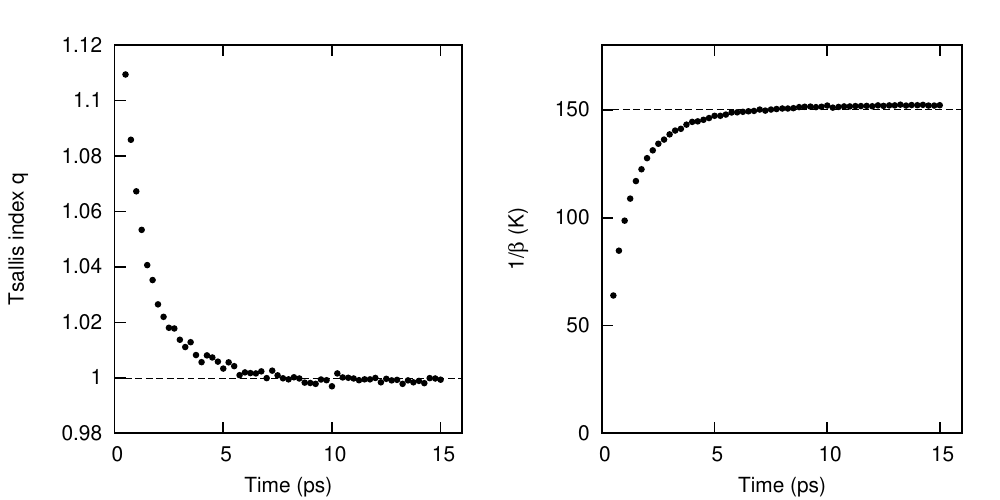}
   \end{center} 
  \caption{\label{fig:2} Relaxation of Tsallis distributions  to
    Maxwell-Boltzmann. Tsallis entropic index $q$ (left) and the parameter $1/\beta$ 
    (right)  vs. time,  for  $\veps=150$ K and $N=216$.}
\end{figure}

The histograms were built in the following way. One starts with a
choice of an initial point, fixing the initial positions of the ions
at their equilibrium positions, while their velocities are extracted
from a Maxwellian at a suitable temperature which ensures that a
prefixed specific energy $\veps$ is obtained (the value of temperature
has to be equal to $2\veps$).  This implies that one has a
Maxwell-Boltzmann distribution for the mode energies.  Then the
equations of motion are integrated using the Verlet method with an
integration step of 2 femtoseconds, up a time of the order of some
nanoseconds.  Then one chooses a time increment $t$ and a sequence of
times $t_n=n\delta t$, and computes the energy-changes $\Big |
E_{\vett k,\rho}(t_n+t) -E_{\vett k,\rho}(t_n)\Big |$.  The time
interval $\delta t$ has to be taken sufficiently large that the
sequence of positions $\vett x_{j,s}(t_n)$ may be considered to produce a
sample representative of the statistics.  Thus the time interval has
to be increased as the specific energy $\veps$ is decreased: we took
$\delta t=5$ picoseconds for specific energies larger than 300 K,
increasing it up to $\delta t=100$ picoseconds at a specific energy of
10 K. Such a procedure was repeated for a number of 16 up to 48
different initial data.

In such a way a sample of the energy-changes was produced, and one can
pass to build the histogram of their moduli. The result, for a system
with a cell of
$N=216$ particles and specific energy $\veps=150$ K, is shown in
Fig.~\ref{fig:1} in log-log scale: the left panel refers to $t=0.5$
picoseconds, and the right one to $t=15$ picoseconds. The sample is
constituted of $ 5\cdot 10^6$ energy-changes.  In the same figure are
plotted both the curve corresponding to the Maxwell-Boltzmann
distribution (dashed line) and the curve corresponding to a Tsallis
distribution (continuous line) with $q=1.11$ e $1/\beta=63.8$ K for
the left panel, and $q=0.9995$ and $1/\beta=152,1$ K for the right
panel. One notices that in the left panel the histogram clearly
differs from Maxwell-Boltzmann, being remarkably well described by a
Tsallis distribution.  However, after a sufficient lapse of time the
histograms, Maxwell-Boltzmann and Tsallis, do coincide: the system did
relax. In all the different tests carried out, with several specific
energies $\veps$ and at several times $t$, the Tsallis distribution was
always found to give a very good description of the histograms. As already
said,  the
reason of such a fact remains at the moment unknown.

\begin{figure}
  \includegraphics[width=\textwidth]{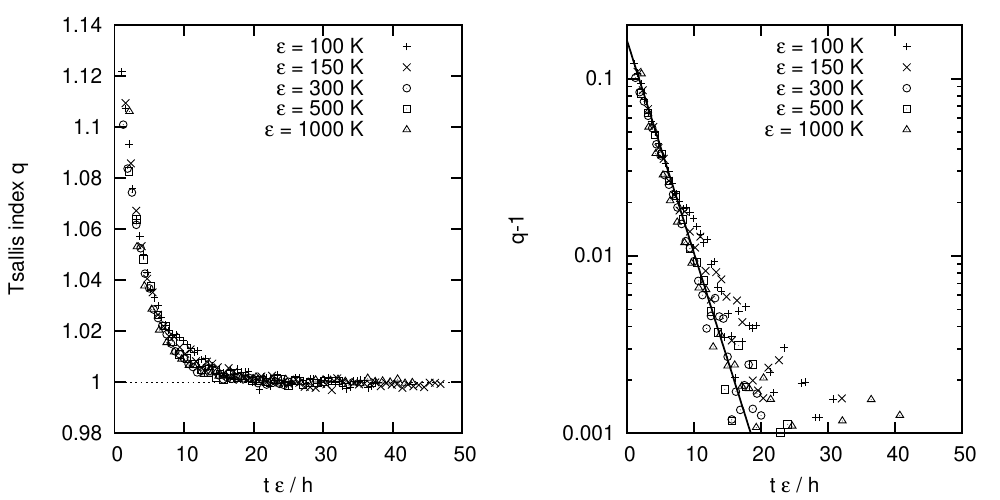}
  \caption{\label{fig:3} Tsallis entropic index $q$ versus rescaled
    time $t \veps$ (in units of $h$), for specific energies $\veps$
    between 150 and 1000 K, in linear scale (left) and in
    semilogarithmic scale (right). This is a main result of the paper,
    since all graphs are seen to superpose.}
\end{figure}

In order to study the relaxation times one can report the plots of the
Tsallis entropic index $q$ versus time. This is shown in
figure~\ref{fig:2}, for  specific energy $\veps=150$ K, and
$N=216$, where the relaxation to Maxwell-Boltzmann is clearly
exhibited.  The computation can be repeated for other values of
specific energy, checking whether one has convergence or not.  The
surprising result obtained is that all plots corresponding to $\veps\ge 150$ K
actually superpose if the entropic index $q$ is plotted as a function
of $\veps t$, or better of $\veps t/h$, which is a dimensionless
variable.\footnote{The choice of Planck's constant $h$ may appear
  natural as the characteristic action of atomic physics. However, in our case
  the choice was induced by the fact that, having estimated the relaxation
  time $\tau$ for several values of $\veps$, we found $\tau(\veps)
  \simeq C/\veps$ with a certain action $C$, and it turned out that
  one has $C\simeq 3h$.}  This is shown in figure~\ref{fig:3} (left
panel). One visually ``sees'' that $q$ attains the value $1$ in a dimensionless
time of about $10$. A less arbitrary characterization
of the relaxation times
can be obtained by
referring to the right panel of the same figure, in which the entropic
index $q$ is plotted as a function of $\veps t/h$ in semilogarithmic
scale. As one can check, the decay of $q$ to 1 follows with very good
approximation a straight line with a slope of $0.28$, which gives a
characteristic dimensionless time of $3.5$.  Recalling that for a
Maxwell-Boltzmann distribution the standard deviation $\Delta E$ is
equal to mean energy $\veps$, the relation between relaxation time and
energy fluctuation takes the form $\tau \cdot\Delta E \simeq \hbar$, which
reminds of the analogous uncertainty relation of quantum mechanics.
\begin{figure} 
   \includegraphics[width=\textwidth]{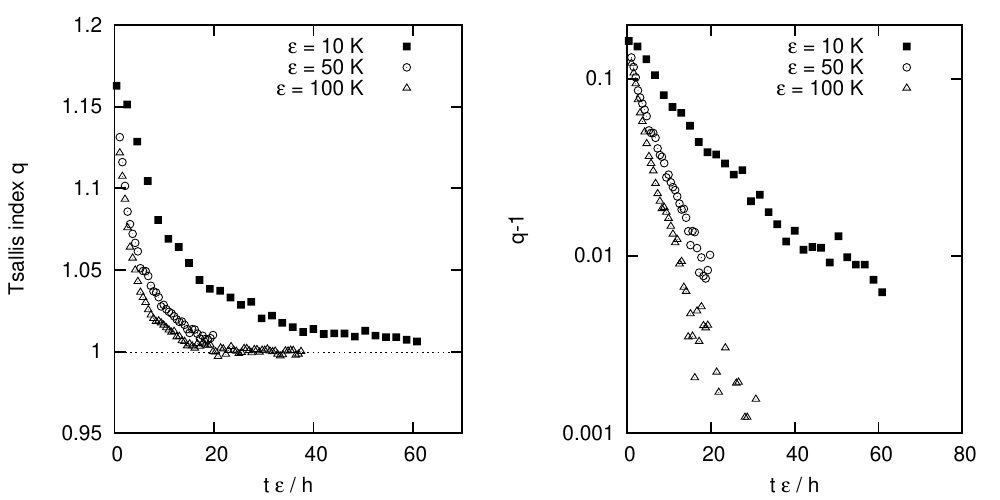}
   \caption{\label{fig:4} Same as Figure~\ref{fig:3}, for low specific
     energies, namely $\veps=$10, 50 and 100 K. Superposition no more
     occurs, through the previous rescaling.}
\end{figure}

\begin{figure}
  \begin{center}
    \includegraphics[width=0.6\textwidth]{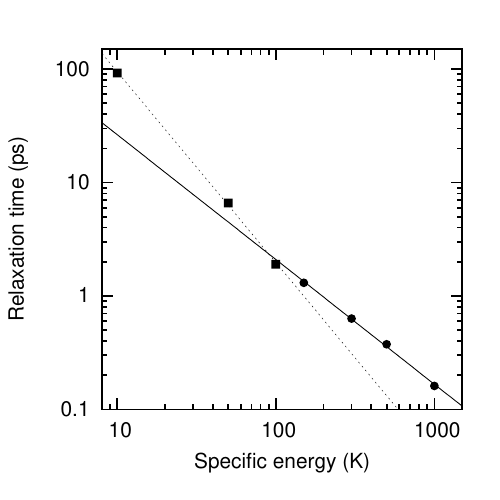}
  \end{center}
  \caption{\label{fig:5} Relaxation time vs specific energy in
    log--log scale, exhibiting a crossover at about 100 K. Here
    $N$=216.}
\end{figure}

For smaller specific energies the rescaling previously illustrated
doesn't apply any more, as is seen from figure~\ref{fig:4}. In fact,
the graphs of $q$ versus the rescaled time $t\veps/h$,
drawn for $\veps=$ 10 and 50 K,
no more superpose each other, neither do they superpose with the
previous ones (here drawn for $\veps =100$).
The right panel shows that the relaxation is at any
rate well described, for any $\veps$, by an exponential,
so that one can  determine the corresponding
relaxation time. The values found are reported versus specific energy,
in log--log scale, in figure~\ref{fig:5}. One sees that a sharp knee
appears at $\veps\simeq$ 100 K: this might signal a change in the
dynamics of the system.

\begin{figure}
  \begin{center}
    \includegraphics[width=\textwidth]{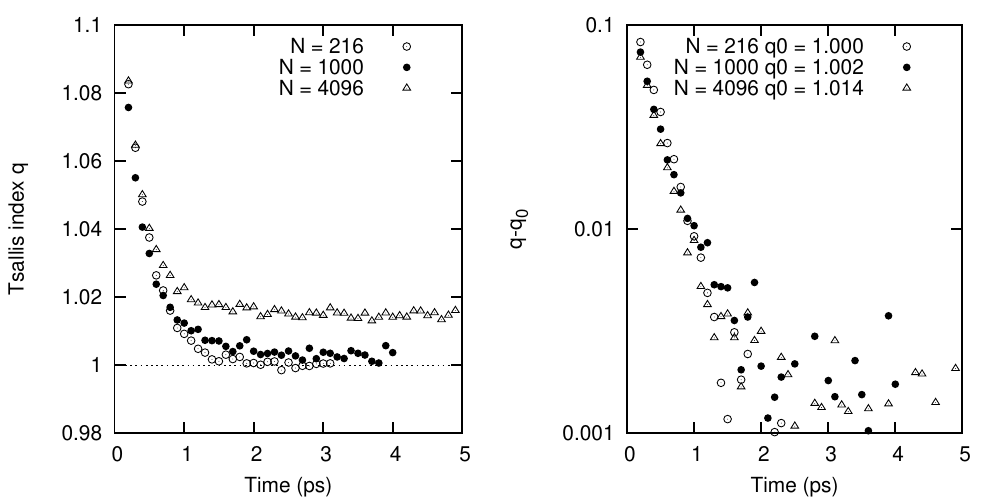}
  \end{center}
  \caption{\label{fig:6} Dependence of the Tsallis entropic index $q$
    on the number of ions $N$ in a cell, at $\veps=500$ K. The result
    is given for $N=$ 216, 1000 and 4096 ions. Left panel: plot of $q$
    vs. time, showing the dependence of the asymptotic value $q_0$ on
    the number of ions. Right panel: plot of the difference betwen the
    entropic index $q$ and its asymptotic value $q_0$ vs. time, in
    semilogarithmic scale. The different plots superpose, showing that
    the relaxation time does not depend on $N$. }
\end{figure}

All the results exhibited in the previous figures were obtained for a
system with a cell of $N=216$ ions. An obvious question is then what
happens by increasing the number of ions.  As we are dealing with a
three-dimensional model involving long range imteractions, increasing
the number of ions implies a more than linear growth of the
computational times, so that we were not in a position to perform a
systematic study. Thus we limited ourselves to raising the number of
ions for just one fixed value of specific energy, actually $\veps=500$
K. The computations were performed at $N=$ 216, 1000 and 4096, and in
all cases it turned out that the statistics of the fluctuations
follows very well a Tsallis distribution. The entropic parameter $q$
obviously depends on time, and the three graphs of $q$ versus time are
reported in figure~\ref{fig:6}.

In particular, in the left panel the plot of $q$ vs. time is reported in
linear scale:
it is apparent that one has  a relaxation but, at least on the time
scale investigated, the asymptotic value $q_0$ of  $q$ 
depends on the number of ions.  We are unable to say whether, at
a larger time, one will finally get Maxwell-Boltzmann, but at any rate
one can conclude that the time for attaining Maxwell-Boltzmann  would grow
for sure when the number of degrees of freedom is increased. This is
at variance with what occurs for Fermi-Pasta-Ulam models with next
neighbors interactions, and is instead in agreement with the results
found for systems with long range interactions. In particular, it is
coherent with the findings of paper \cite{christodoulidis14}, which deals with a
one-dimensional Fermi-Pasta-Ulam model with long range interactions:
the entropic index relaxes to a metastable value, which lasts for a
longer time as the number of particles is increased, until eventually
attaining the Maxwell-Boltzmann value $q=1$.  By the way, for values
of $q$ close to 1, our computations have a large margin of error, so
that it seems to be premature to advance precise statements.

The right panel reports instead the entropic index $q$ minus the
asymptotic value $q_0$ vs. time, in semi logarithmic scale, for
different values of $N$: one sees that the points superpose well along
a straight line. This implies that the relaxation to $q_0$ is
exponential, and that the relaxation time doesn't depend on $N$, at
least at the specific energy $\veps=500$ K considered.

\section{Discussion of the results}

Let us now summarize and comment the results obtained for the moduli
of the energy-changes occurring with initial data of generic type (i.e., with
a Maxwell-Boltzmann distribution for the mode energies), for the
Lithium Fluoride model.  First of all, the moduli of the energy-changes follow
time-dependent Tsallis distributions which present a relaxation after
a rather well defined relaxation time $\tau(\veps)$.  Concerning the
nature of the distribution after relaxation, we don't yet have a
definitive answer, but the result obtained for 4096 ions seems not to
exclude, as a possible conjecture, that for a large enough number $N$
of ions, and perhaps even at the thermodynamic limit, the distribution
still be a Tsallis one. If this feature were confirmed, one would be
in presence of a relevant fact, since it would mean that classical
statistical mechanics shouldn't be the standard Gibbs one.

However, interesting phenomena already are found even for small
numbers of ions, with the occurence of quantum-like features involving a
characteristic action of the order of Planck's constant, within a wholly
classical frame. Such features concern the form of the curve $\tau (\veps)$ of
relaxation time versus specific energy, which indeed was shown to be
pretty well defined in terms of the results available for a small
value $N$ of ions.

The first phenomenon, exhibited in figure~\ref{fig:5}, is the
existence of a threshold (or a knee, or a crossover) $\veps_0$,
apparently at 100 K, i.e., rather near to the value of 120 K that would
be expected from the results on the infrared spectra.\footnote{Our
  impression is that such a discrepancy should be due to the fact that
  the results on the spectra were obtained at fixed pressure, whereas
  the results presented here were obtained at fixed volume.}  One may
wonder which role be played by such a threshold.  We already recalled
how, for the threshold observed in the standard Fermi-Pasta-Ulam,
already in the early years 70's  Cercignani put forward the suggestive
hypothesis that it be the analogue of zero-point energy.  That
hypothesis actually produced a strong stimulus on the subsequent works
on the Fermi--Pasta--Ulam problem, in view of its relevance for the
relations between classical and quantum mechanics.  Such a perspective
appeared to have been frustrated when a general consensus was obtained
on the fact that in systems of Fermi-Pasta-Ulam type
equipartition is eventually attained, also
starting from initial data faraway from it, since such a fact was
interpreted as confirming the common opinion that classical mechanics
fails at low temperatures. Things started changing, however, with the
investigations on the realistic Lithium Fluoride model which, at
variance with the standard Fermi-Pasta-Uam model, allowed one to
compare numerical computations with experimental data, for example in
connection with polaritons (see ref.~\cite{CGLS2014}) and with the
infrared spectra.

In the present paper a further comparison, leading support to the
Cercignani hypothesis, is made possible, since one finds that the
threshold $\veps_0$ agrees with the quantum  zero-point energy, albeit up to a
factor 2.  The simplest way to see this fact, is to proceed as
Einstein did in his work on specific heats, namely, by assimilating
the crystal to a system of oscillators all of the same frequency
$\nu$, equal to the characteristic infrared frequency of the
crystal. In the Lithium Fluoride case this means the spectral line at
300 cm$^{-1}$ (i.e., $\nu= 300\ c$ cm$^{-1}$, where $c$ is the speed of
light). In fact, recalling that one has $hc/k_B =1.4$, if one assumes
$\veps_0= 100$ K one immediately gets
\begin{equation}
  \frac 12 h \nu = 2\veps_0 \ ,
\end{equation}
i.e., $\veps_0$ is just one half the quantum value $h\nu/2$.
Essentially the same result is obtained also if one sums the
zero-point energies of all the single normal modes of the model.

The second phenomenon, in addition to the first one on the existence
of the threshold $\veps_0$, is the law for the relaxation time
observed above it, which, as was previously pointed out, reminds of the
time-energy uncertainty relation of quantum mechanics, and still
exhibits a characteristic
action, very close to Planck's constant.

So the problem arises of understanding how is it possible at all that
Planck's constant shows up in a purely classical system.  Now, it
would seem obvious that Planck's constant appears in our model through
the coefficients of the phenomenological Buckingham potential, which
takes implicitly into accout the forces among ions due to the
electrons, in the adiabatic hypothesis. Thus the occurring of Planck's
constant would eventually be a manifestation of the quantum character
of the dynamics of the electrons. In such a way one may understand why
Planck's constant shows up not only in the zero-point energy, but also
above the knee, in the rescaling of the relaxation times through
specific energy $\veps$.  More in general, the occurring of a
characteristic action of the order of Planck's constant through the
effective potential might explain why classical dynamics may reproduce
the experimental result, as it actually does so well in our model, for
the infrared spectra.

In fact, however, the situation is more complex, as exhibited typically
by the result obtained in the paper \cite{CGGG2020b}, concerned with
the $H_2^+$ ion.  Such an ion is the simplest case in which an effective
potential due to electrons occurs, since one has just one electron
which produces a chemical bond between two protons. In such a case, an
effective potential can be computed in quantum mechanical terms, thus
involving Planck's constant in an obvious way.  However, in the
mentioned paper we showed how in the $H_2^+$ ion the binding through
an effective potential is explained also in classical terms in the
three-body purely Coulomb model, which involves no free parameter at
all.  The relevant point is that the binding occurs\footnote{As is
  well known, in systems with pure Coulomb forces the equilibrium points
  are always unstable. In our case, instead, the bond is produced by
  the attractive action of the electron, which continues to  bounce
  back and forth
  between the protons, and turns out to  prevail on the repulsion
  between the two protons.}  only if the motion of the system is
sufficiently regular
rather than chaotic; otherwise no confinement can occur, which is
indeed the core of the adiabatic principle in a classical frame. In
particular it is found that the binding occurs only if the angular
momentum of the electron about the axis through the protons is
sufficiently large, actually of the order $h$.  So, in the case of the
$H_2^+$ ion, Planck's constant enters as a chaoticity threshold, in a
purely classical frame. Perhaps, an analogous situation may be found
to occur also for the phenomenological inter-ionic potential in the
general case involving more than just one electron.

\section{Conclusions}

The numerical investigation performed in the present work does not
allow one to answer in a conclusive way the main question addressed in
this paper.  Namely, whether the necessity of introducing
temperature values different from those of specific energy in order to
reproduce the susceptibility data at low tempertatures, originates from
the intrinsically quantum character of the dynamics, or rather from a
lack of sufficient chaoticity in the classical dynamics. Indeed,
Figure~\ref{fig:6} shows that the results depend on the number $N$ of
ions constituting the system. This dependence is very strong, because
the exponent in the Tsallis distribution , i.e., $1/(1-q)$, changes from
$\infty$, to  500 and  70
by just multiplying by 4 the particles' number $N$ (with the asymptotic
value $q_0$ passing from 1 to 1.002 and 1.014),
so that is not clear what occurs at the thermodynamic limit.

This notwithstanding, a fixed point seems to have been established,
namely, that the Tsallis distribution plays a relevant role for our
system and more in general, we believe, for ionic crystals.  Which
might be the dynamical properties (apart from a reduced chaoticity)
inducing the occurrence of such distributions, is not clear. Neither
are clear the implications such distributions may have for the
expected values of the thermodynamic variables. By the way, one might
recall in this connection that, as previously mentioned,
the relevant quantities in Statisitcal
Thermodynamics actually are not the energies themselves, which follow
Maxwell-Boltzmann distributions, but rather, as particularly stressed
 by Nernst \cite{nernst}, the \emph{exchanged}
energies, i.e., the energy-changes, whose moduli were here shown to follow
Tsallis distributions. In any case, certainly chaoticity diminishes
for decreasing specific energy, as witnessed by the occurring of the
knee in the graph of relaxation time versus specific energy. It would
be interesting to establish what happens as the number of ions is
increased, namely, whether the marked tendency for a diminishing of
chaoticity with increasing $N$, exhibited at the higher temperatures,
persists also below the threshold $\veps_0$. For our program,
addressed at explaining the phenomenology of Lithium Fluoride in
purely classical terms, this is is crucial point, on which we hope to
come back in the near future.

As a final remark, we would like to point out that we are well aware
of the difficulties of such a program; but it is not excluded that it
may be implemented. In such a case, the relations between classical
and quantum mechanics would appear under a new light, with quantum
mechanical features occurring as a manifestation of lack of full
chaoticity. The present paper, however, shows at least that the
relations between classical and quantum mechanics are more complex
than is commonly believed.

\vspace{2.em}
\noindent
\textbf{Acknowledgements}:
  Research carried out with the support of
  resources of Big and Open Data Innovation Laboratory (BODaI-Lab),
  University of Brescia, granted by Fondazione Cariplo and Regione
  Lombardia.

\bibliography{biblio}
\bibliographystyle{unsrt}

\end{document}